\documentclass[12pt]{article}
\usepackage{graphicx,color,epsfig,rotate}
\title{
Ground states of the generalized Falicov-Kimball model
in one and two dimensions} 
\author{Pavol Farka\v sovsk\'y and Hana \v{C}en\v{c}arikov\'a \\
Institute  of  Experimental  Physics,  Slovak   Academy   of
Sciences\\
Watsonova 47, 043 53 Ko\v {s}ice, Slovakia}
\date{}
\begin{document}
\baselineskip=20pt
\maketitle

\begin{abstract}
A combination of small-cluster exact-diagonalization  calculations 
and a well-controlled approximative method is used to study 
the ground-state phase diagram of the spin-one-half
Falicov-Kimball model extended by the spin-dependent on-site interaction
between localized ($f$) and itinerant ($d$) electrons. Both the magnetic
and charge ordering are analysed as  functions of the spin-dependent 
on-site interaction ($J$) and the total number of itinerant ($N_d$) and 
localized ($N_f$) electrons at selected $U$ (the spin-independent 
interaction between the $f$ and $d$ electrons).  
It is shown that the spin-dependent interaction (for $N_f=L$,  
where $L$ is the number of lattice sites) stabilizes 
the ferromagnetic (F) and ferrimagnetic (FI) state, while the stability 
region of the antiferromagnetic (AF) phase is gradually reduced. The precisely 
opposite effect on the stability of F, FI and AF phases has a reduction 
of $N_f$. Moreover, the strong coupling between the $f$ and $d$-electron 
subsystems is found for both $N_f=L$ as well as  $N_f < L$.
\end{abstract}
\thanks{PACS nrs.:75.10.Lp, 71.27.+a, 71.28.+d, 71.30.+h}

\newpage
\section{Introduction}
In the past decade, a considerable amount of effort has been devoted to
understand the underlying physics that leads to a charge
ordering in strongly correlated electron systems. The motivation was
clearly due to the observation of a such ordering in doped 
nickelate~\cite{Ni} and cuprate~\cite{Cu} materials, some of which 
constitute materials that exhibit high-temperature superconductivity.
One of the simplest models suitable to describe charge ordered
phases in interacting electron systems is the Falicov-Kimball
model (FKM)~\cite{Falicov}. Indeed, it was shown that already the 
simplest version of this model (the spinless FKM) exhibits an
extremely rich spectrum  of charge ordered solutions, including various
types of periodic, phase-separated and striped phases~\cite{Lemanski1}. 
However, the spinless version of the FKM, although non-trivial, is not able
to account for all aspects of real experiments. For example, 
many experiments show that a charge superstructure
is accompanied by a magnetic superstructure~\cite{Ni,Cu}.
In order to describe both types of ordering in the unified picture
Lemanski~\cite{Lemanski2} proposed a simple model based on a 
generalization of the spin-one-half FKM 
with an anisotropic, spin-dependent local interaction that couples 
the localized and itinerant subsystems. The model Hamiltonian is

\begin{equation}
H=\sum_{ij\sigma}t_{ij}d^+_{i\sigma}d_{j\sigma}+
U\sum_{i\sigma\sigma'}f^+_{i\sigma}f_{i\sigma}d^+_{i\sigma'}d_{i\sigma'}+
J\sum_{i\sigma\sigma'}(f^+_{i-\sigma}f_{i-\sigma}-f^+_{i\sigma}f_{i\sigma})
d^+_{i\sigma}d_{i\sigma},
\end{equation}
where $f^+_{i\sigma}$, $f_{i\sigma}$ are the creation and annihilation
operators for an electron of spin $\sigma=\uparrow,\downarrow$ in the 
localized state at lattice site $i$ and $d^+_{i\sigma}$, $d_{i\sigma}$ 
are the creation and annihilation operators of the itinerant electrons 
in the $d$-band Wannier state at site~$i$.

The first term of (1) is the kinetic energy corresponding to
quantum-mechanical hopping of the itinerant $d$ electrons
between sites $i$ and $j$. These intersite hopping
transitions are described by the matrix  elements $t_{ij}$,
which are $-t$ if $i$ and $j$ are the nearest neighbors and
zero otherwise (in the following all parameters are measured
in units of $t$).
The second term represents the on-site
Coulomb interaction between the $d$-band electrons with density
$n_d=N_d/L=\frac{1}{L}\sum_{i\sigma}d^+_{i\sigma}d_{i\sigma}$
and the localized $f$ electrons with density
$n_f=N_f/L=\frac{1}{L}\sum_{i\sigma}f^+_{i\sigma}f_{i\sigma}$,
where $L$ is the number of lattice sites. The third term 
is the above mentioned anisotropic, spin-dependent (of the Ising type)
local interaction between the localized and itinerant electrons
that reflects the Hund's rule force. 
Moreover, it is assumed that the on-site Coulomb interaction 
between $f$ electrons is infinite and so the double 
occupancy of $f$ orbitals is forbidden. 

Thus from the major interaction terms that come into account for the
interacting $d$ and $f$ electron subsystems only the Hubbard type 
interaction between the spin-up and spin-down $d$ electrons has been omitted
in the Hamiltonian (1). In his work~\cite{Lemanski2} Lemanski presents
a simple justification for the omission of this term, based on 
an intuitive argument: the longer time electrons occupy the same 
site, the more important becomes interaction between them. According 
to this rule the interaction between the itinerant $d$ electrons 
$(U_{dd})$ is smaller than the interaction between the localized
$f$ electrons $(U_{ff})$ as well as smaller than the spin-independent 
interaction between the localized and itinerant electrons $U$. In this paper 
we specify more precisely conditions when this term can be neglected.
For this reason we start our study with the case $U_{dd} \ne 0$.
To determine the effects of $U_{dd}$ interaction on the ground-states
of the conventional spin-one-half FKM ($J=0$) the exhaustive studies
of the ground-state phase diagrams of the model (in the $n_f-U_{dd}$
plane) are performed for several cluster sizes. Of course, an
inclusion of the $U_{dd}$ term makes the Hamiltonian (1) intractable
by methods used for the conventional spin-one-half/spinless FKM
and thus it is necessary to use other numerical methods.
Here we use the Lanczos method to study exactly the ground states
of the spin-one-half FKM generalized with $U_{dd}$ interaction
between the spin-up and spin-down $d$ electrons.

\section{Results and discussion}
\subsection{The spin-1/2 FKM with the Hubbard interaction between 
itinerant electrons} 

The Hamiltonian of the spin-1/2 FKM with the Hubbard interaction between 
itinerant $d$ electrons can be written as the sum of three terms: 
\begin{equation}
H=\sum_{ij\sigma}t_{ij}d^+_{i\sigma}d_{j\sigma}+
U\sum_{i\sigma\sigma'}f^+_{i\sigma}f_{i\sigma}d^+_{i\sigma'}d_{i\sigma'}+
U_{dd}\sum_{i}d^+_{i\uparrow}d_{i\uparrow}d^+_{i\downarrow}d_{i\downarrow}.
\end{equation}
Since the $f$-electron density operators
$f^+_{i\sigma}f_{i\sigma}$ of each site $i$ commute with the 
Hamiltonian (2), the $f$-electron occupation number is a good 
quantum number, taking only two values, $w_{i\sigma}=0,1$
according to whether the site $i$ is unoccupied or occupied
by the localized $f$ electron of spin $\sigma$.
Therefore the Hamiltonian (2) can be rewritten as
\begin{equation}
H=\sum_{ij\sigma}h_{ij}d^+_{i\sigma}d_{j\sigma}+
U_{dd}\sum_{i}d^+_{i\uparrow}d_{i\uparrow}d^+_{i\downarrow}d_{i\downarrow},
\end{equation}
where $h_{ij}=t_{ij}+U(w_{i\downarrow}+w_{i\uparrow} )\delta_{ij}$.
This Hamiltonian can be considered as a generalized Hubbard model. 
To determine the ground-state energy of the model we have used the Lanczos
method~\cite{Dag}. However, since the hopping amplitudes depend now on the 
$f$-electron distribution $w=\{w_1,w_2 \dots w_L\}$ 
(we remember that the double occupancy of $f$ orbitals 
is forbidden $w_i=w_{i\uparrow}+w_{i\downarrow}=0,1$)
the Lanczos procedure has to be used many times (strictly said
$L!/((L-N_f)!N_f!)$ times) for given $L$ and $N_f$.
Of course, such a procedure demands in practice a considerable amount of 
CPU time, which imposes severe restrictions on the size of clusters that 
can be studied within the exact-diagonalization method. For this reason 
we were able to investigate exactly only the clusters up to $L=12$.  
Fortunately, it was found that in some parameter regimes the ground-state
characteristics of the model are practically independent of $L$ and thus
already such small clusters can be used satisfactorily to represent the 
behavior of macroscopic systems. In particular, we have studied the stability
of the ground-state configuration $w^0(N_f)$ (obtained for $U_{dd}=0$ and 
fixed $N_f$) at finite values of $U_{dd}$. The results of numerical 
calculations obtained for $U=4$ and $U=8$ are summarized in Fig.~1 in the 
form of $n_f-U_{dd}$ phase diagrams (the half-filled band case $n_f+n_d=1$
is considered). One can see that the ground-state 
configuration $w^0(N_f)$ found for $U_{dd}=0$ persists as a ground state 
up to relatively large values of $U_{dd}$ ($U^c_{dd}\sim 2$, for $U=4$ 
and $U^c_{dd}\sim 6$, for $U=8$), revealing small effects of the 
$U_{dd}$ term on the ground state of the model in the strong $U$ 
interaction limit. Contrary to the strong coupling case, 
for small ($U=1$) and intermediate ($U=2$) values
of the Coulomb interaction between the localized and itinerant electrons
very strong effects of $U_{dd}$ term on the ground states of the model 
have been observed.  In these cases the typical values of $U^c_{dd}$
are of order 0.5 but for some $N_f$ even much smaller 
values were found. Thus we can conclude that the Hubbard type 
interaction between the spin-up and spin-down $d$ electrons can be 
neglected in the strong interaction limit between the localized
and itinerant electrons ($U \geq 4$). For this reason all next 
calculations on the spin-one-half FKM with spin-dependent 
Coulomb interaction $J$ between the $f$ and $d$ electrons have 
been done at $U=4$.

\subsection{The spin-1/2 FKM with the spin-dependent interaction between 
itinerant and localized electrons} 

The spin-dependent interaction term $H_J$ between the itinerant ($d$) 
and localized ($f$) electrons does not violate the condition 
$[f^+_{i\sigma}f_{i\sigma},H]=0$ and thus the Hamiltonian (1) can 
be rewritten as

\begin{equation}
H=\sum_{ij\sigma}h_{ij}d^+_{i\sigma}d_{j\sigma},
\end{equation}
where $h_{ij}=t_{ij}+(Uw_i+Jw_{i\downarrow}-Jw_{i\uparrow})\delta_{ij}$.
Thus for a given $f$-electron configuration
$w=\{w_1,w_2 \dots w_L\}$, the Hamiltonian (4)
is the second-quantized version of the single-particle
Hamiltonian $h(w)$, so the investigation of the model (4) 
is reduced to the investigation of the spectrum of $h$ for different 
configurations of $f$ electrons. This can be performed exactly, over 
the full set of $f$-electron distributions (including their spins), 
or approximatively, over the reduced set of $f$-electron
configurations. The second way has been used in the original work
by Lemanski~\cite{Lemanski2}. He studied the two-dimensional version of 
the model using the method of restricted phase diagrams (all possible
configurations of the localized $f$ electrons for which the number
of sites per unit cell is less or equal to 4 are considered) and 
presented some preliminary results concerning the charge and magnetic
order in the ground-state of this model. For example, he detected 
various phases with complex charge and magnetic structures that form
consecutive stages of transformation of F to AF
phase with an increase of the band filling. In the present work we 
study the one (D=1) and two (D=2) dimensional analogue of the model. To 
determine the ground states of the model we use the method of small
cluster-exact diagonalization calculations in a combination with a 
well-controlled numerical method~\cite{Fark1}. 
   
Since the $d$ electrons do not interact among themselves, the
exact numerical calculations on finite clusters precede directly 
in the following steps:
(i) Having $U,J$ and $w=\{w_1,w_2 \dots w_L\}$ fixed, find
all eigenvalues $\lambda_k$ of $h(w)$. (ii) For a given
$N_f=\sum_iw_i$ and $N_d$ determine the ground-state energy
$E(w)=\sum_{k=1}^{N_d}\lambda_k$ of a particular
$f$-electron configuration $w$ by filling in the lowest
$N_d$ one-electron levels (the spin degeneracy must be taken
into account). (iii) Find the $w^0$ for which
$E(w)$ has a minimum. Repeating this procedure for different
values of $N_f,N_d,U$ and $J$, one can immediately study
the ground-state phase-diagrams of the model in different 
parameter spaces.

To reveal an influence of the anisotropic, spin-dependent interaction
between the localized and itinerant electrons on the ground states of the
model we have started the study with the case $N_f=L$ (D=1). In this case
each lattice site is occupied by one (up or down) $f$ electron 
(the double occupancy is forbidden) and thus only distributions over 
different spin configurations should be examined. Although the total number 
of configurations increases very rapidly with the cluster size $L$ (as
$2^L$), relatively large lattices can be reached by this method ($L\sim32$)  
if all symmetries of the Hamiltonian are considered. In Fig.~2 we summarize
numerical results obtained by small-cluster exact diagonalization
calculations on the largest cluster that we were able to consider
exactly ($L=32$) in the form of $N_d-J$ phase diagram. To avoid an ambiguity
in determination of FI and AF phases
we examined the ground states only for even $N_d$. In the $J$ direction the 
calculations have been done with step $\Delta J = 0.05$. Various 
phases that enter into the phase diagram are classified according to 
$S^z_f=\sum_iw^0_{i\uparrow}-w^0_{i\downarrow}$ and 
$S^z_d=N_{d\uparrow}-N_{d\downarrow}$: the ferromagnetic phase 
is characterized by $|S^z_f|=N_f, |S^z_d|=N_d$, the ferrimagnetic phases 
are characterized by $0 < |S^z_f| < N_f, 0 < |S^z_d| < N_d$ and 
the antiferromagnetic phases  are characterized by $|S^z_f|=0, 
|S^z_d|=0$.  Comparing numerical results obtained for $|S^z_f|$ 
and $|S^z_d|$ one can find a nice correspondence between the
magnetic phase diagrams of localized ($f$) and itinerant ($d$) subsystems. 
Indeed, with the exception of several isolated points at $J=0.05$, 
the corresponding F, FI and AF phases perfectly coincide
over the remaining part of diagrams showing on the strong coupling 
between the magnetic subsystems of localized and itinerant electrons 
for nonzero values of $J$. It is interesting that already very small
changes of the spin-dependent interaction can produce so important 
cooperative changes. This confirms the supposition that the 
spin-dependent interaction between the localized and itinerant electrons
could play an important role in description of ground state 
properties of the generalized FKM. In general, the spin-dependent 
interaction $J$ stabilizes the F and FI
phases while the AF phase is gradually suppressed
with increasing $J$. Moreover, we have observed that within 
the AF phase (with the exception of cases 
$N_d=14,26$) the ground states (for given $N_d$) do not change 
when $J$ increases, while within the FI phase
very strong effects of $J$ on ground states were found (see Fig.~2).
For example, the transition from the AF to 
F phase at $N_f=12$ realizes through the following
sequence of FI phases: 
$[\uparrow_2\downarrow_6]_4 \to
\uparrow_{21}\downarrow_2\uparrow_2\downarrow_3\downarrow_2
\uparrow_2 \to
\uparrow_{26}\downarrow_2\uparrow_2\downarrow_2 \to
\uparrow_{30}\downarrow_2$, 
where the lower index denotes the number of consecutive sites occupied 
by up or down spin $f$ electrons, or the number 
of repetitions of the block $[\dots]$.
In Fig.~2 we present also the complete set of 
ground-state  configurations (obtained on the above specified set
of $N_d$ and $J$ values) from the AF region.
Between them one can find different types of periodic and non-periodic 
configurations, but the most interesting examples represent 
configurations formed by antiparallel F domains, that 
illustrate convincingly the cooperative effects of 
spin-dependent interaction $J$ between the localized and itinerant 
electrons.   

Let us turn our attention to the case $N_f\ne L$. From the numerical point 
of view this case is considerably exacting, since now we have to minimize
the ground state energy not only over all different spin configurations
but also over all different $f$-electron distributions. 
This takes a considerable amount of CPU time and for this reason 
we were able to investigate exactly only the clusters up to $L=24$ for 
$N_f\ne L$. In Fig.~3 we present the one-dimensional skeleton phase diagram 
of the generalized FKM in the $N_f-N_d$ plane obtained by small-cluster exact
diagonalization calculations for $L=24, U=4$ and $J=0.5$.
Again, the stability regions of AF, F and FI phases are marked by ($\cdot$), 
(+) and ($\circ$). Here we displayed the numerical results 
only for the localized subsystem since the analysis of $|S^z_d|$ and 
$|S^z_f|$ showed that the F, FI, and AF phases 
corresponding to localized and itinerant subsystems coincide also for 
$N_f < L$ (similarly as for $N_f=L$, only a few exceptions have been 
observed in isolated points that result probably from the finite-size 
effects). The most striking feature of the $N_f-N_d$ phase diagram 
is that with decreasing $N_f$ the AF phase is stabilized,
while the F and FI phases are suppressed. It is 
interesting that this effect is strongly asymmetric and a 
disappearance of F and FI phases realizes in two
different ways. For small $d$-electron concentrations the F 
phase disappears practically immediately outside the point
$N_f=L$. The FI phase survives along the main diagonal 
in the narrow band and fully disappears near the point 
$N_f=L/2, N_d=L$. In the opposite limit 
(large $d$-electron concentrations) the F phase persists
for a wide interval of $N_f< L$ and with decreasing $N_f$ 
disappears gradually. The same behaviour exhibits also the FI phase.    
The F phase appears also for $N_f\to 0$. However, in this limit 
strong finite-size effects have been observed, and thus it is 
questionable if this phase persists really in the thermodynamic limit. 

Although the skeleton phase diagram of the generalized FKM is rather
simple, the spectrum of magnetic solutions that yields the model for
the AF and FI phases is very rich. Indeed,
for $L=24$ and $J=0.5$ we have found 140 different AF
phases and 20 different FI phases that enters into the 
$N_f-N_d$ phase diagram. Of course, it is not possible to present
here all ground-state configurations, but let us show
at least several main configurations types, with the largest
stability regions.  
From the AF phases the largest stability region (denoted by I) corresponds 
to configurations of the type 
$\uparrow_n\downarrow_n0_{L-2n}$. 
The second largest region (denoted by II) corresponds
to AF configurations of the type 
$[\uparrow0_n\downarrow0_n]_k0_{L-2k(n+1)}$.
Typical examples of the AF ground states from the 
central region of the phase diagram represent periodic configurations 
of the type 
$\uparrow_n0_{m}\downarrow_n0_{m}$
(below the main diagonal)
and configurations of the type
$\uparrow_2[\downarrow\uparrow]_{k_1}\downarrow_2
0_{m}[\downarrow0_p\uparrow0_p]_{k_2}$, or 
$\uparrow_2[\downarrow\uparrow]_{k_1}\downarrow_2
0_{m}[\downarrow0_p\uparrow0_{p-1}]_{k_2}$,
above the main diagonal. 
Between these configurations and the F region
the ground states are the segregated configurations of the type
$\uparrow_2[\downarrow\uparrow]_k\downarrow_20_{m}$, or
$[\uparrow\downarrow]_{k_1}\uparrow_2[\downarrow\uparrow]_{k_2}\downarrow_2
[\uparrow\downarrow]_{k_3}0_{m}$, or their modifications (the region denoted 
by III).
In the FI region the typical examples of ground states
represent configurations of the type
$\uparrow_n[0\downarrow0\uparrow]_k0\downarrow0$.

These results show that the spectrum of magnetic and charge 
solutions that yields the spin-one-half FKM model generalized
with the spin-dependent interaction between the localized and itinerant
electrons is indeed very rich. Of course, one can ask if these
phases persist also on larger clusters. Unfortunately, lattices
larger than $L=32$ (for $N_f=L$), or $L=24$ (for $N_f < L$)
are beyond the reach of present day computers within the 
exact diagonalization technique. Therefore, to resolve this 
problem one has to use other numerical methods. Very promising
seems to be the well-controlled numerical method that we have elaborated 
recently to study ground states of the spinless FKM~\cite{Fark1}. 
This method is described in detail in our previous
papers~\cite{Fark1,Cenci}
and thus we summarize here only the main steps of the algorithm
that is a simple modification of the exact-diagonalization
algorithm described above:
(i) Chose a trial configuration $w=\{w_1,w_2, \dots ,w_L\}$.
(ii) Having $w$, $U$ and $J$ fixed, find
all eigenvalues $\lambda_k$ of $h(w)$. (iii) For a given
$N_f=\sum_iw_i$  and $N_d$ determine the ground-state energy
$E(w)=\sum_{k=1}^{N_d}\lambda_k$ of a particular
$f$-electron configuration $w$ by filling in the lowest
$N_d$ one-electron levels.
(iv) Generate a new configuration $w'$ by moving a randomly
chosen electron to a new position which is chosen also at random.
(v) Calculate the ground-state energy $E(w')$. If $E(w')<E(w)$
the new configuration is accepted, otherwise $w'$ is rejected.
Then the steps (ii)-(v) are repeated until the convergence
(for given parameters of the model) is reached.
Of course, one can move instead of one electron (in step (iv))
simultaneously two or more electrons, thereby the convergence of the method 
is improved. Indeed, tests that we have performed for a wide range 
of the model parameters showed that the latter implementation of the method, 
in which  $1 < p < p_{max}$ electrons ($p$ should be chosen at random) are 
moved to new positions overcomes better the local minima of the ground state 
energy. In this paper we perform calculations with $p_{max}=N_f$. The main 
advantage of this implementation is that in any iteration step the system
has a chance to lower its energy (even if it is in a local minimum), 
thereby the problem of local minima is strongly reduced (in principle, the
method becomes exact if the number of iteration steps goes to infinity). On 
the other hand a disadvantage of this selection is that the method converges 
slower than for $p_{max}=2$ and $p_{max}=3$. To speed up the convergence of 
the method (for $p_{max}=N_f$) and still to hold its advantage we generate 
instead the random number $p$ (in step (iv)) the pseudo-random number $p$ 
that probability of choosing decreases (according to the power law) with 
increasing $p$. Such a modification improves considerably the convergence 
of the method.

To test the convergence of the method we have first calculated 
the ground-state configurations of the model in the $N_f-N_d$ plane 
on the cluster of $L=24$ sites. Comparing these results with exact ones
(discussed above) we have found that the method is able to reproduce
the exact ground states after relatively small number of iterations
(typically 5000-10000 iterations per site).   
Then we have used the method to study the $N_f-N_d$ phase diagram 
of the model on larger clusters consisting of $L=36$
and $L=48$ sites. Our numerical computations 
showed that all main results obtained on small clusters
hold also on larger clusters. Again we have observed strong 
coupling between two magnetic subsystems and a coincidence
of corresponding magnetic phases, that stability regions
are practically unchanged with increasing $L$ (see Fig.~4). 
Moreover, we have observed that the main configurations types 
found for $L=24$ persist also on large clusters and thus we
suppose that the real magnetic phase diagram of the model 
will be very close to ones presented in Fig.~3 and Fig.~4,  
Of course, it is possible that some ground states   
that are uniform on the finite lattices could be degenerated
in the thermodynamic limit.

The same calculations we have performed also in two dimensions.
The two-dimensional results are of particular importance 
since they could shed light on the mechanism of two-dimensional
charge and magnetic ordering in doped nickelate~\cite{Ni} and 
cuprate~\cite{Cu} materials. Here we concern our attention
on a description of basic types of charge and magnetic ordering 
that exhibits the spin-one-half FKM with spin-dependent 
interaction between $d$ and $f$ electrons in two dimensions.
To minimize the finite-size effects the numerical calculations
have been done on three different clusters of $4\times 4, 6\times 6$ 
and $8\times 8$ sites. On the $4\times 4$ cluster the calculations 
were performed by exact-diagonalization method and on larger clusters 
the approximative method described above was used. 

Similarly as in the one dimension we start our two-dimensional studies
with the case $N_f=L$. The magnetic phase diagrams of the $f$ and $d$
electron subsystems calculated for $N_f=L$ ($L=36$) 
are shown in Fig.~5. Comparing these phase diagrams with their
one-dimensional counterparts one can find obvious similarities.  In both
cases the basic structure of the phase diagram is formed by three 
large F, FI and AF domains that are accompanied by secondary phases.
However, while in the one dimension the secondary phases are stable only 
in isolated points at very small values of $J$, in two dimensions
these secondary phases persist also for large $J$. Calculations that 
we have performed on different clusters ($4 \times 4, 6 \times 6$, and  
$8 \times 8$) showed that the secondary structure depends very strongly on 
the cluster-size (with increasing $L$ it is gradually suppressed) and 
it is not excluded that it fully disappears in the thermodynamic limit
$(L\to \infty)$.

The typical examples of the ground-state configurations (that represent
the most frequently appearing types of the ground states in the $N_d-J$
phase diagram) are displayed in Fig.~6.  Again one can see that 
the spectrum of magnetic solutions that yields the FKM extended by
spin-dependent interaction is very rich. In addition to the F phase 
(that the stability region shifts to higher $d$-electron concentrantions
when $J$ increases) there are various types of AF and FI
structures like the antiparallel F domains (2-3),
the axial magnetic stripes (4-7),
the diagonal magnetic stripes (8-11)
and the perturbed diagonal magnetic stripes (12-15). This again 
demonstrates strong effects of the spin-dependent interaction
on the formation of magnetic superstructures in the extended FKM
and its importance for a correct description of correlated electron 
systems.

From the experimental point of view the most interesting case is,
however, the case $N_f<L$ that could model the real situation 
in doped nickelate and cuprate systems~\cite{Ni,Cu}.
To describe possible charge and magnetic orderings for $N_f<L$
we have performed an exhaustive studies of the model on 
the $6\times6$ and $8\times8$ clusters.  Numerical calculations
have been done over the full set of $N_f$ and $N_d$ values
and they revealed a rich spectrum of coexisting charge and magnetic 
superstructures. Some typical examples of these superstructures      
are presented in Fig.~7.
Between them one can find various types of phase segregated (e.g., 1), 
phase separated (e.g., 16) and $n$-molecular (e.g., 3) configurations 
with F,FI and AF ground states as well as various types of axial 
(e.g., 9) and diagonal (e.g., 5) magnetic/charge
stripes. In generally, we have observed that the system has 
tendency towards phase segregation for small and large $d$-electron
concentrations, while near the $n_d=1$ point the system prefers    
to form the various types of axial and diagonal stripes. 
In addition, similarly as in $D=1$ a strong reduction 
of $F$ and $FI$ phases with decreasing $N_f$ is observed also in $D=2$. 
We have found that the same tendencies and the same types of 
configurations persist on both examined latices ($6\times 6$ and 
$8\times 8$), confirming the stability of obtained results. 

Although we have presented here only the basic types of charge and 
magnetic superstructures (a more complete set will be given elsewhere)
they clearly demonstrate an ability of the model to describe different
types of charge and magnetic ordering. This opens an alternative route for 
understanding  of formation an inhomogeneous charge/magnetic order 
in strongly correlated electron systems.
In comparison to previous studies of this phenomenon based
on the Hubbard and $t-J$ model~\cite{stripes}, the study within the 
generalized spin-one-half FKM has one essential advantage 
and namely that it can be performed in a controllable way
(due to the condition $[f^+_{i\sigma}f_{i\sigma},H]=0$),
and in addition it allows easily to incorporate and examine
effects of various factors (e.g., an external magnetic field,
nonlocal interactions, etc.) on formation of charge and magnetic
superstructures. The work in this direction is currently in progress.

In summary, a combination of small-cluster exact-diagonalization 
calculations and a well-controlled approximative method was used to study 
the ground-state phase diagrams of the generalized
spin-1/2 FKM with an anisotropic, spin-dependent 
on-site interaction between localized and itinerant electrons
for $N_f=L$ ($0 \leq J \leq 1$ in $D=1$ and $0 \leq J \leq 4$ in $D=2$) 
and $N_f < L$ ($J=0.5$).  
For both cases it was observed that the anisotropic, spin-dependent
interaction induces strong coupling between the localized and itinerant 
subsystems and that the magnetic phase diagrams of these 
subsystems coincide. In general, the spin-dependent interaction 
(for $N_f=L$) stabilizes the F and FI state, while  
the stability region of AF phase is gradually reduced.
The opposite effect on the F, FI and AF phases
has a reduction of $N_f$. For both $N_f=L$ and $N_f < L$ an 
extremely rich spectrum of charge and magnetic solutions has been found.
In particular, we have observed various types of phase segregated, 
phase separated, striped, periodic and nonperiodic charge/spin distributions 
that clearly demonstrate strong cooperative effects of spin-dependent
interaction on the ground states of the model.

\vspace{0.5cm}
This work was supported by the Slovak Grant Agency VEGA
under grant No. 2/4060/04 and the Science and Technology Assistance
Agency under Grant APVT-20-021602. Numerical results were obtained using
computational resources of the Computing Centre of the Slovak 
Academy of Sciences.

\newpage
\begin{figure}[hb]
\includegraphics[angle=0,width=14cm,scale=1]{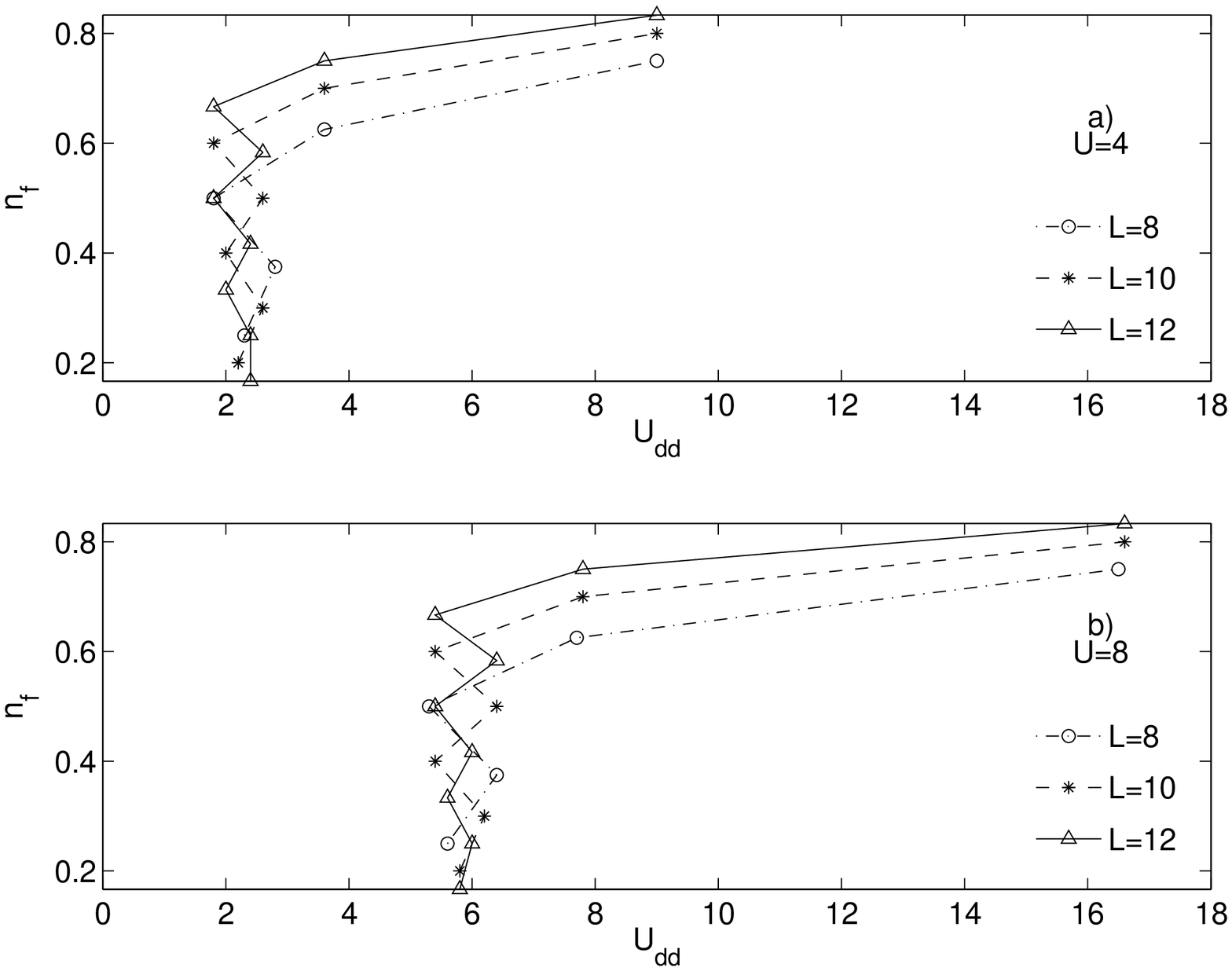}
\caption{ 
The ground-state phase diagrams of the spin-one-half FKM 
extended by the Hubbard interaction between the itinerant electrons
calculated for two different values of $U$
on small finite clusters of $L=8,10$ and 12 sites.
Below $U^c_{dd}$ the ground states are the ground-state
configurations of the conventional spin-one-half FKM ($U_{dd}=0$).
Above $U^c_{dd}$ these ground states become unstable.
The one-dimensional exact-diagonalization results.
}
\label{fig1}
\end{figure}

\newpage
\begin{figure}[hb]
\begin{center}
\includegraphics[angle=0,width=14cm,scale=1]{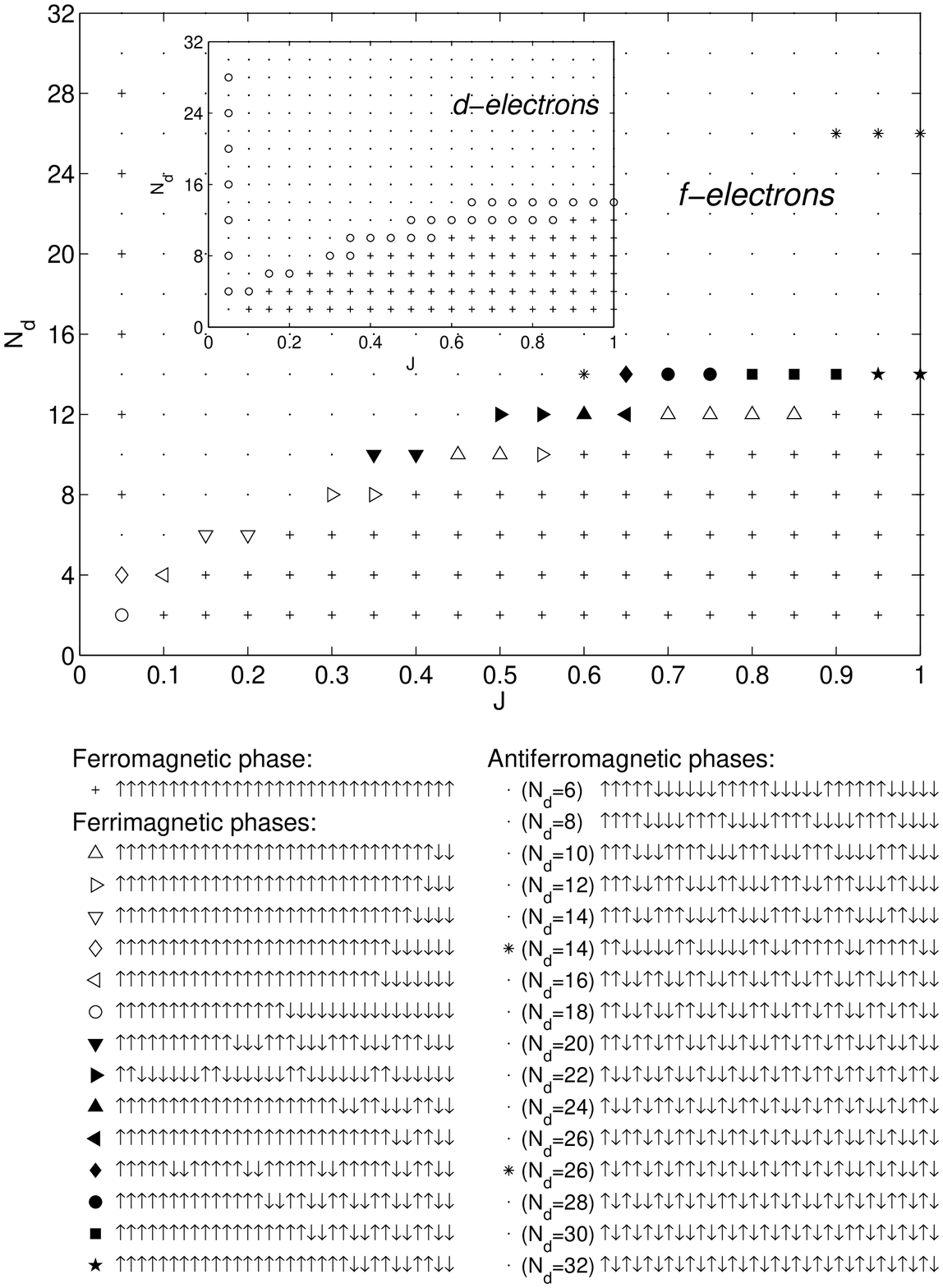}
\end{center}
\caption{ 
The $f$-electron ground-state phase diagram of the spin-one-half FKM 
extended by the spin-dependent interaction  calculated for $N_f=L$ ($L=32$).
The one-dimensional exact-diagonalization results.
Inset: 
The $d$-electron ground-state phase diagram of the model
calculated at the same values of model parameters.
($\cdot$): the AF phase, (+): the F phase, ($\circ$): the FI phase.
}
\label{fig2}
\end{figure}

\newpage
\begin{figure}[hb]
\begin{center}
\includegraphics[angle=-90,width=14cm,scale=1]{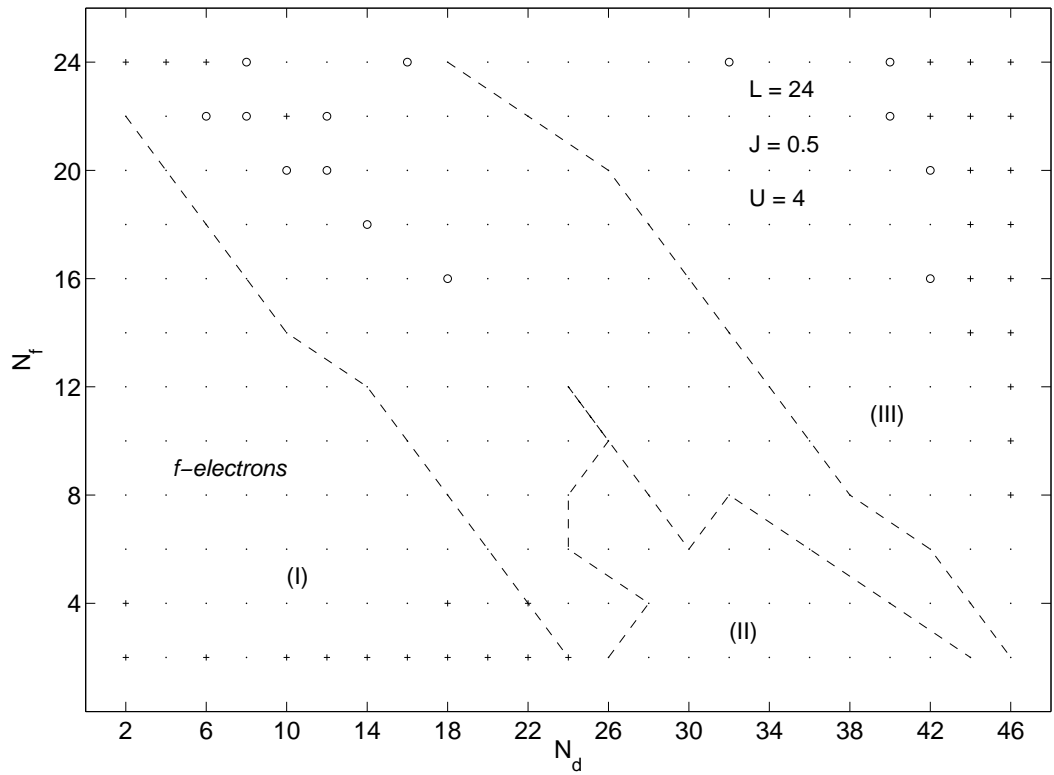}
\end{center}
\caption{ 
The $f$ electron skeleton phase diagram of the spin-one-half FKM 
extended by the spin-dependent interaction  calculated for $N_f\leq L, J=0.5$ 
and $L=24$. ($\cdot$): the AF phase, (+): the F phase, ($\circ$): the FI phase.
The one-dimensional exact-diagonalization results.
}
\label{fig3}
\end{figure}

\newpage
\begin{figure}[hb]
\begin{center}
\includegraphics[angle=0,width=13.5cm,scale=1]{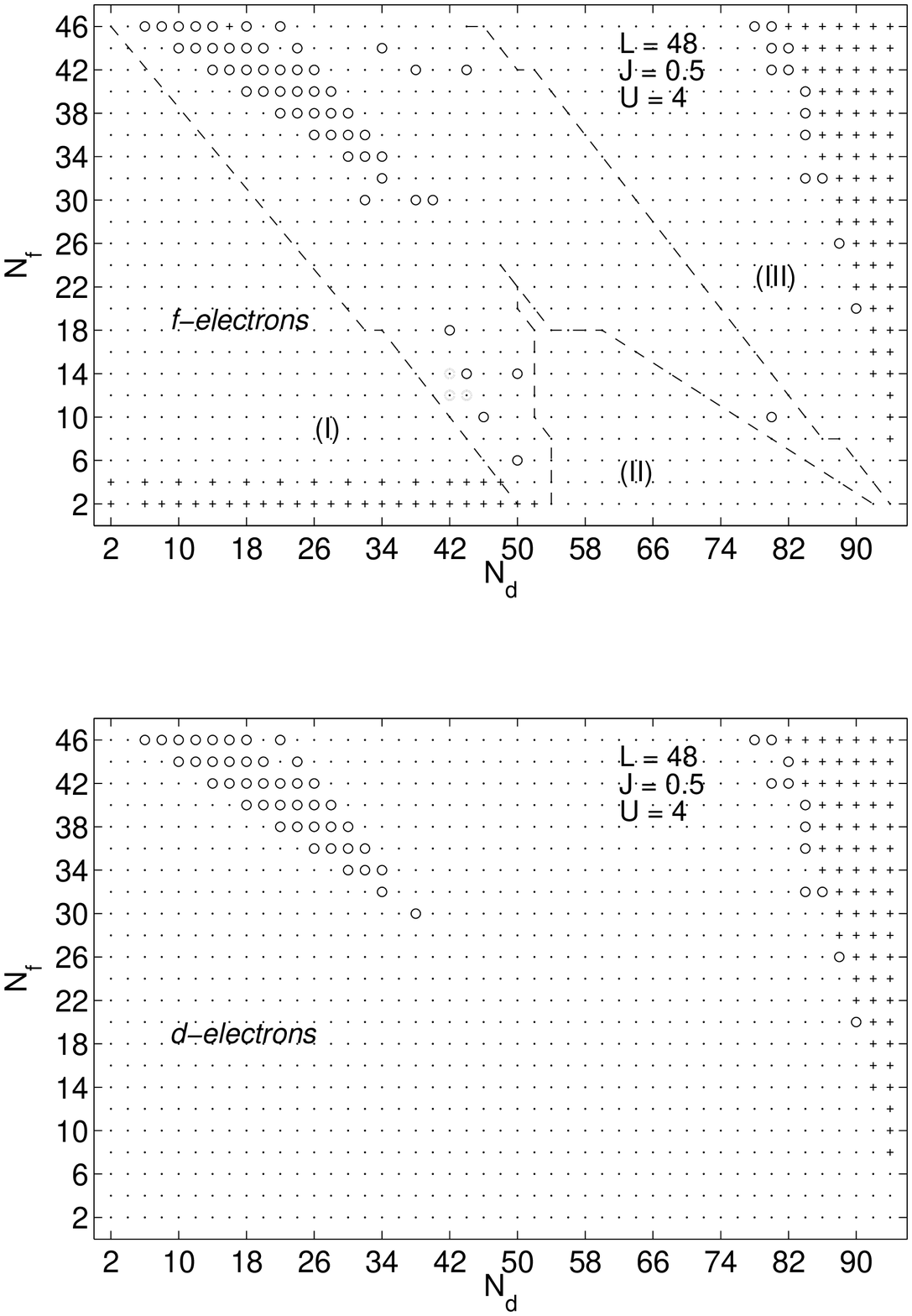}
\end{center}
\caption{ 
The $f$ and $d$ electron phase diagrams of the spin-one-half FKM 
extended by the spin-dependent interaction  calculated for $N_f < L, J=0.5$ 
and $L=48$.
($\cdot$): the AF phase, (+): the F phase, ($\circ$): the FI phase.
The one-dimensional approximative results.
}
\label{fig4}
\end{figure}

\newpage
\begin{figure}[hb]
\begin{center}
\includegraphics[angle=0,width=14cm,scale=1]{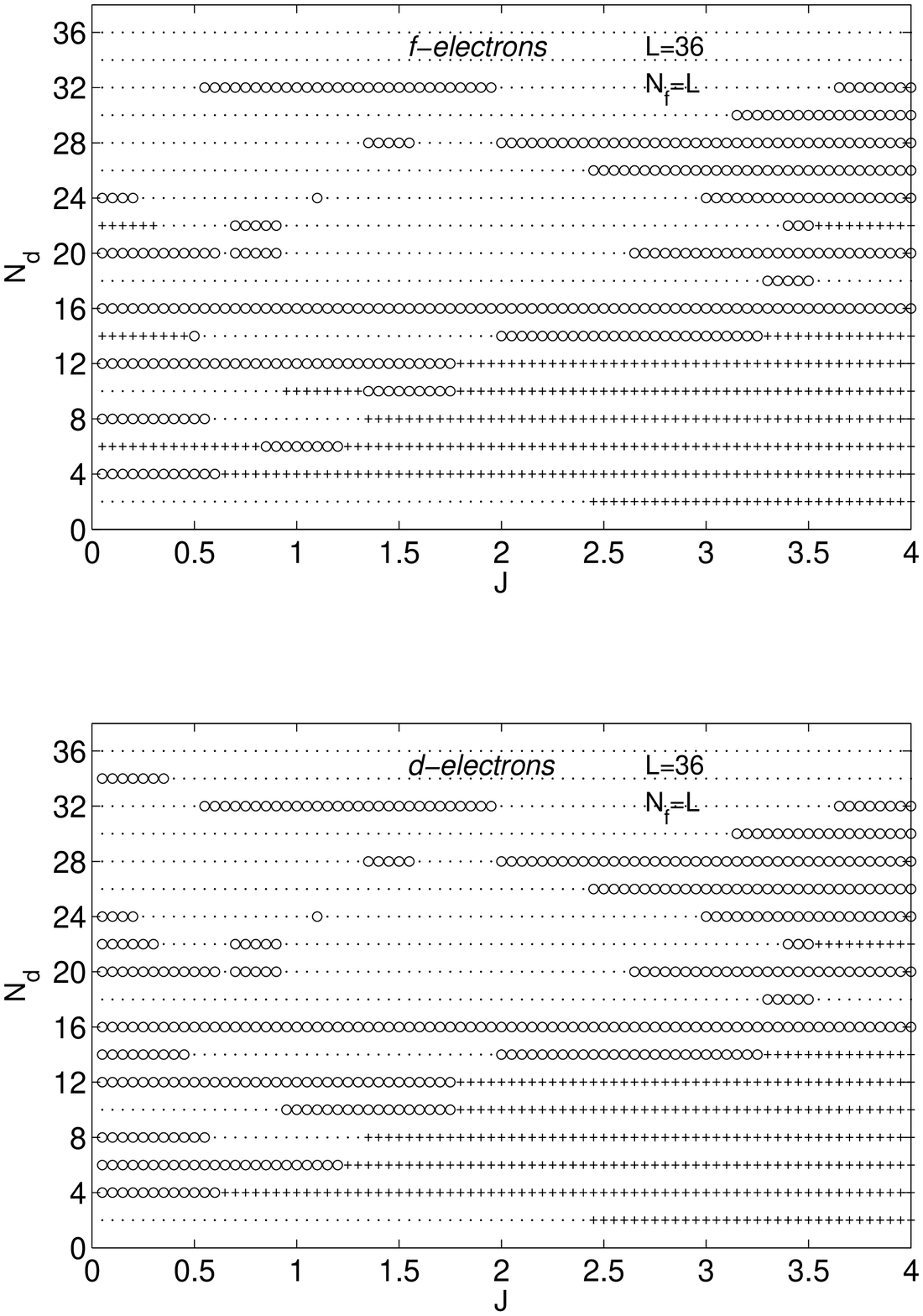}
\end{center}
\caption{ 
The $f$ and $d$ electron ground-state phase diagrams of the spin-one-half FKM 
extended by the spin-dependent interaction  calculated for $N_f=L$ ($L=36$).
The two-dimensional approximative results.
}
\label{fig5}
\end{figure}

\newpage
\begin{figure}[hb]
\begin{center}
\includegraphics[angle=0,width=11.5cm,scale=1]{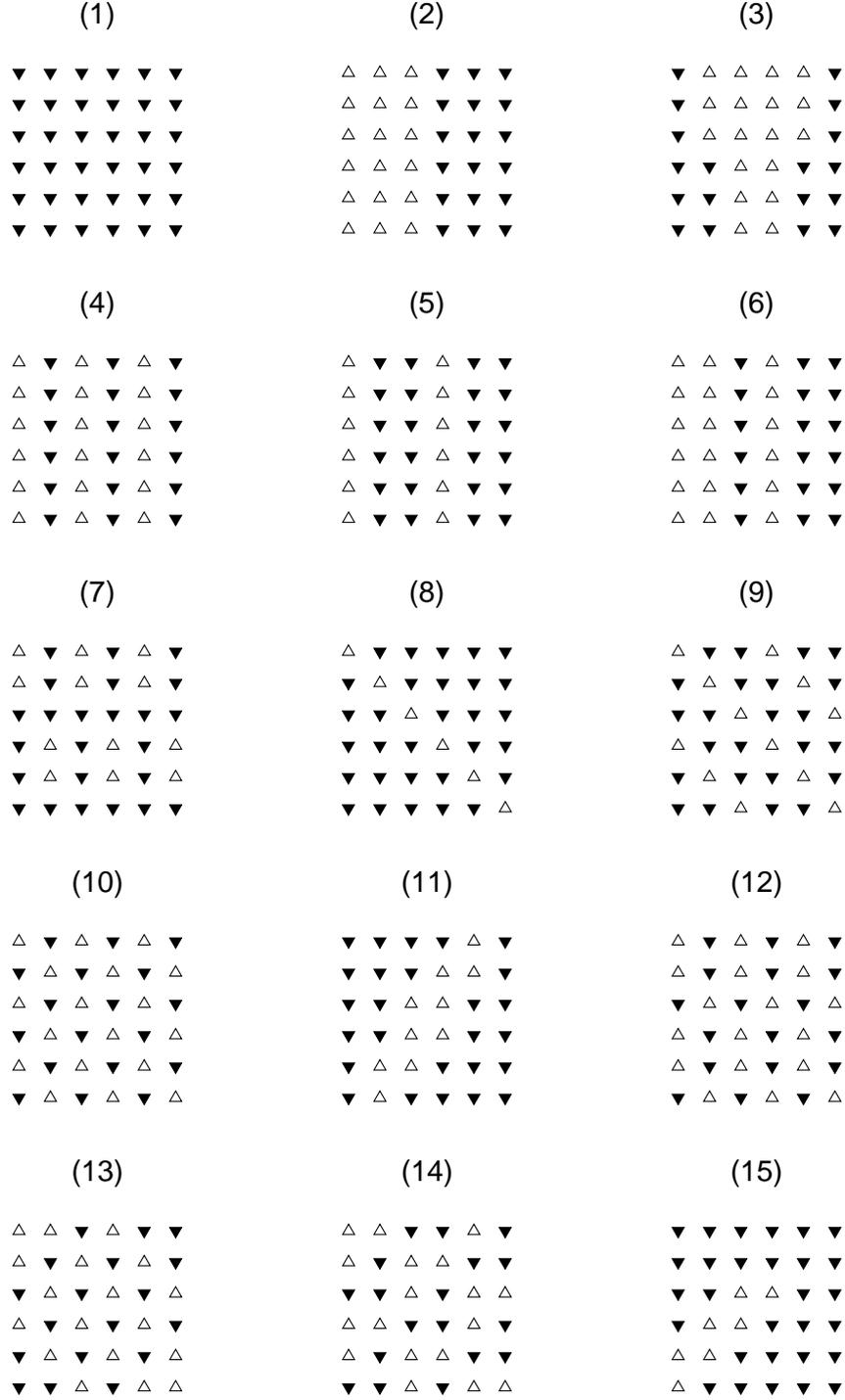}
\end{center}
\caption{ 
Typical examples of ground states of the spin-one-half FKM 
extended by the spin-dependent interaction  obtained for $N_f=L$ ($L=36$). 
To visualise the spin distributions we use $\bigtriangleup$ 
for the up spin electrons and $\bigtriangledown$
for the down spin electrons. 
The two-dimensional approximative results.
}
\label{fig6}
\end{figure}

\newpage
\begin{figure}[hb]
\begin{center}
\includegraphics[angle=0,width=13cm,scale=1]{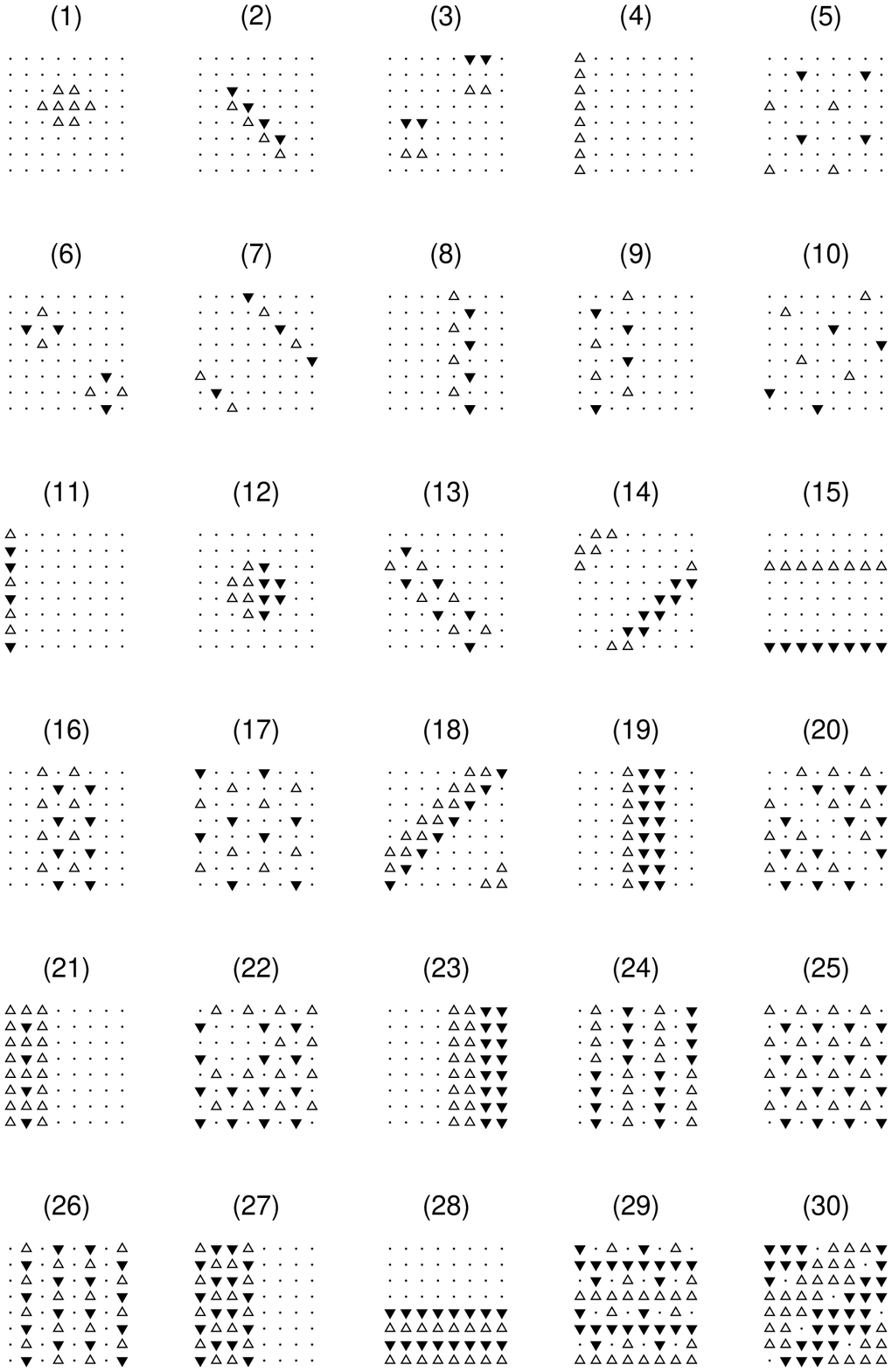}
\end{center}
\caption{ 
Typical examples of ground states of the spin-one-half FKM 
extended by the spin-dependent interaction  obtained for $N_f < L,J=0.5$ and
$L=64$.
The two-dimensional approximative results.
}
\label{fig7}
\end{figure}


\newpage

\begin{thebibliography}{999}
\bibitem{Ni} C. H. Chen, S.-W. Cheong and A. S. Cooper, Phys. Rev. Lett.
{\bf 71}, 2461 (1993); J. M. Tranquada, D. J. Buttrey, V. Sachan and J. E.
Lorenzo, Phys. Rev. Lett.~{\bf 73}, 1003 (1994); Phys. Rev. B~{\bf 52}, 
3581 (1995); V. Sachan {\it et al.}, {\it ibid.} {\bf 51}, 12742 (1995). 

\bibitem{Cu} J. M. Tranquada, B. J.
Sternlieb, J. D. Axe, Y. Nakamura and S. Uchida, Nature (London) {\bf 375}, 
561 (1995); Phys. Rev. B~{\bf 54}, 7489 (1996); Phys. Rev. Lett.~{\bf 78}, 
338 (1997); H. A. Mook, P. Dai and F. Dogan, Phys. Rev. Lett.~{\bf 88}, 
097004 (2002); 

\bibitem{Falicov} L.M. Falicov and J.C. Kimball, Phys. Rev. Lett.
{\bf 22}, 997 (1969).

\bibitem{Lemanski1} R. Lemanski, J. K. Freericks and G. Banach, Phys. Rev. Lett.
{\bf 89}, 196403 (2002);

\bibitem{Lemanski2} R. Lemanski, Phys. Rev. B {\bf 71}, 035107 (2005).

\bibitem{Dag} E. Dagotto, Rev. Mod. Phys. {\bf 66}, 763 (1994).

\bibitem{Fark1} P. Farka\v{s}ovsk\'y, Eur. Phys. J.  B {\bf 20}, 209 (2001).

\bibitem{Cenci} H. \v{C}en\v{c}arikov\'a and P. Farka\v{s}ovsk\'y, 
Int. J. Mod. Phys. B{\bf 18}, 357 (2004).

\bibitem{stripes} V. J. Emery, S. A. Kivelson and H. Q. Lin, Phys. Rev. Lett.
{\bf 64}, 475 (1990);\\ L. P. Pryadko, S. A. Kivelson and D. W. Hone, Phys.
Rev. Lett. {\bf 80}, 5651 (1998); S. R. White and D. J. Scalapino, 
Phys. Rev. Lett. {\bf 80},1272 (1998); {\bf 81}, 3227 (1998); 
Phys. Rev. B {\bf 60}, R753 (1999); {\bf 61}, 6320 (2000);
A. M. Oles, Acta Physica Polonica B {\bf 31}, 2963 (2000); J. Frolich and
D. Uettschi, J. Stat. Phys. {\bf 118}, 973 (2005).



\end{thebibliography}
\end{document}